\documentclass[12pt]{iopart}
\usepackage{iopams}
\usepackage{graphicx}
\begin{document}

\title[Electron-induced limitation of plasmon propagation in Ag nanowires]{Electron-induced limitation of surface plasmon propagation in silver nanowires}

\author{M. Song$^1$, A. Thete$^2$, J. Berthelot$^{1,\dag}$, Q. Fu$^3$, D. Zhang$^3$, G. Colas des Francs$^1$, E. Dujardin$^2$
and A. Bouhelier$^1$}
\address{$^1$Laboratoire Interdisciplinaire Carnot de Bourgogne, CNRS UMR
6303, Universit\'e de Bourgogne, 9 Avenue Alain Savary, Dijon,
France}
\address{$^2$CEMES, CNRS UPR 8011, Toulouse, France}
\address{$^3$Department of Optics and Optical Engineering,
University of Science and Technology of China, Hefei, People's
Republic of China}
\address{$^\dag$ Now at
ICFO-Institut de Ci\`encies Fot\`oniques, Castelldefels (Barcelona),
Spain}
 \ead{alexandre.bouhelier@u-bourgogne.fr}
%%%%%%%%%%%%%%%%%%%%%%%%%%%%%%%%%%%%%%%%%%%%%%%%%%%%%%%%%%%%%%%%%%%%%
\begin{abstract}
Plasmonic circuitry is considered as a promising solution-effective
technology for miniaturizing and integrating the next generation of
optical nano-devices. A key element is the shared metal network
between electrical and optical information enabling an efficient
hetero-integration of an electronic control layer and a plasmonic
data link. Here, we investigate to what extend surface plasmons and
current-carrying electrons interfere in such a shared circuitry. By
synchronously recording surface plasmon propagation and electrical
output characteristics of single chemically-synthesized silver
nanowires we determine the limiting factors hindering the
co-propagation of electrical current and surface plasmons in these
nanoscale circuits.

\end{abstract}
%%%%%%%%%%%%%%%%%%%%%%%%%%%%%%%%%%%%%%%%%%%%%%%%%%%%%%%%%%%%%%%%%%%%%
 \submitto{\NT} \maketitle
\section{Introduction}
Plasmonic circuitry~\cite{ozbay06,EbessenPT08} is a unique platform
capable for simultaneously sustaining an optical signal encoded in
the form of a surface plasmon and an electrical current. This
emerging technology has attracted tremendous interest for its
potential to interface electronics and photonics at the
nanoscale~\cite{BarnesN03,AtwaterSA07,GramotnevNP10,DragomanPQE08,SchullerNM10}.
The ability to guide surface plasmons and electrons on the same
physical link may enable a seamless integration of an optical data
processing level within an electronic control
architecture~\cite{Zia06,ParkAPL07,NeutensNP09}.

Among the variety of plasmonic waveguides developed to
date~\cite{WeeberPRB99,GrandidierAPL10,MaierAM01,LiuOPEX05,FedutikPRL07},
quasi-one dimensional structures like silver and gold nanowires are
especially desirable because they have potential for realizing dense
plasmonic and electronic routing networks. In particular,
chemically-synthesized metal nanowires offer a unique medium to
guide and manipulate surface plasmon at the nanoscale with improved
performances~\cite{DicksonJPCB00,KrennPRL05,HalasNP07,ConwayOPEX07}.
Surface plasmon modes can be readily excited in these structures by
different coupling
strategies~\cite{DicksonJPCB00,SirbulyPNAS05,PyaytNN08,YanPNAS09,FangNL11}
and signal-processing functionalities were recently demonstrated
including routing~\cite{ManjavacasNL009,FangNL10,LiIEEE11}, logic
functions~\cite{WeiNL11}, modulation~\cite{LiSmall11} and plasmon
electrical detection~\cite{FalkNP09}. Despite an advanced control of
the surface plasmon field, integration of the platform with an
electronic layer has remained elusive and confined to a
thermo-optical activation of dielectric-loaded surface plasmon
stripe waveguides~\cite{Leosson06,Gosciniak10}. For metal nanowires,
it is not clear to which extent surface plasmon characteristics are
retained under direct-current (DC) biasing operation. This question
is however critical to assess the capability of a plasmonic
circuitry to simultaneously sustain surface plasmon propagation and
an electrical current.

In this paper, we investigate the effect of an electron flow on the
surface plasmon properties when these two information channels are
transported synchronously in chemically-synthesized metal nanowires.
By recording surface plasmon characteristics as a function of the
electrical environment, limiting factors affecting the
co-propagation are experimentally determined.

\section{Electrical characterization}
Pentagonal-twinned silver nanowires were synthesized using a
modified polyol process producing nanowires with section width
comprised between 300~nm and 600~nm~\cite{Zhu08}. The colloidal
solution of nanowires was then drop-casted on a glass substrate
pre-patterned with grid landmarks to precisely coordinate the
position of selected nanowires. Electrical contacts were designed by
electron-beam lithography to connect the extremities of individual
nanowires. A 70~nm-thick gold layer was evaporated to form the
conductive leads. Because the thickness of the electrodes was
smaller than the radius of the nanowires, Au evaporation was
performed at a $\sim 50^\circ$ azimutal angle to insure electrical
continuity between the electrodes and the nanowire. The structures
were electrically biased by a regulated low voltage power supply and
current flow was recorded with a home-made current-to-voltage
converter in a two-point configuration.

\begin{figure}
\includegraphics[width=1\textwidth]{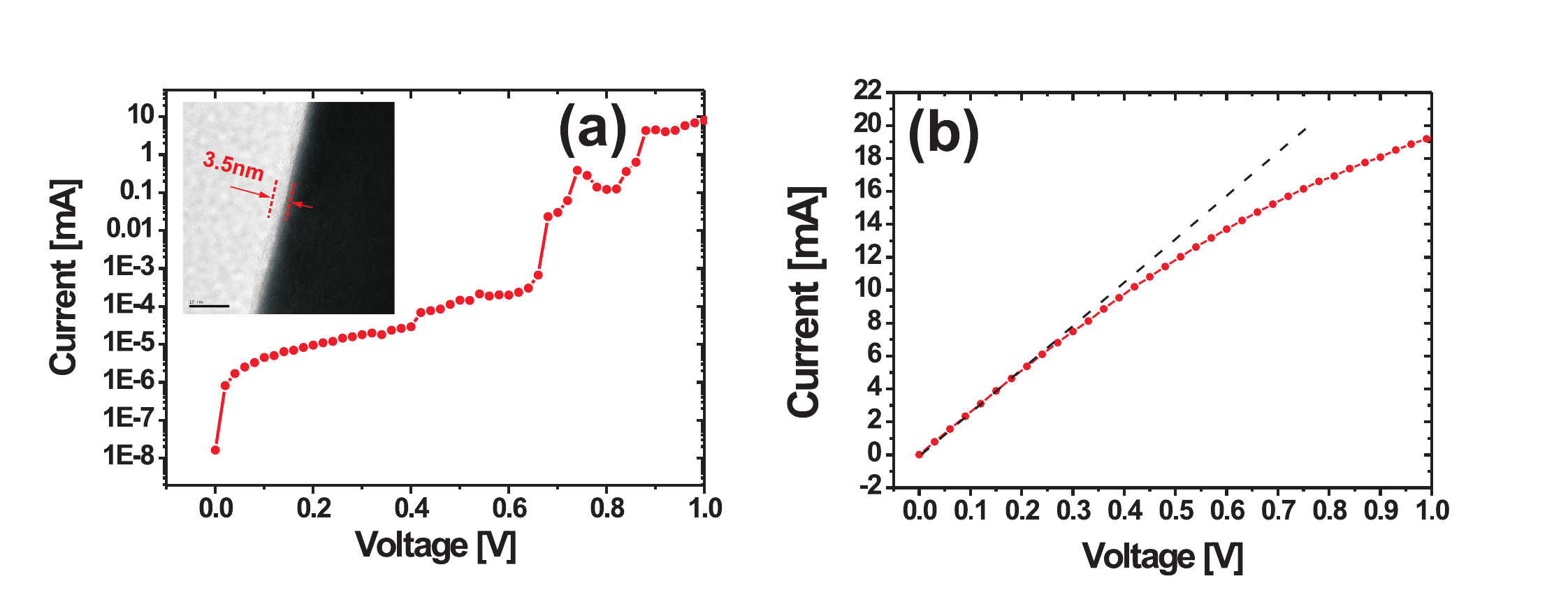}
\caption{(a) Electrical characteristic of a contacted Ag nanowire
during an initial bias sweep. The current flowing through the
nanowire occurs by steps and is very low until 0.8~V. The curve
displayed in semi-logarithmic scale. Inset: TEM image of the PVP
surfactant layer encapsulating a pristine Ag nanowire. (b)
Electrical characteristic of the same nanowire during a subsequent
bias sweep. The curve shows a monotonous evolution of the current
departing from an ohmic behavior after 0.3~V due to a
temperature-dependent resistance.}
 \label{IV}
\end{figure}

A representative electrical output characteristic of a contacted Ag
nanowire is displayed in semi-logarithmic scale in Fig.~\ref{IV}(a)
during an initial bias sweep. Current flowing in the nanowire is
typically low across a large range of bias and occurs by increasing
steps. This non-ohmic current/voltage behavior of the nanowire
originates from the initially poor electrical connection between the
nanowire and the leads. The chemistry employed to synthesize the
nanowires uses an excess of polyvinylpyrrolidone (PVP) surfactant
that is bound to the surface of the nanowire. A transmission
electron micrograph (TEM) showing a $\sim$3.5~nm thick surfactant
layer is displayed in the inset of Fig.~\ref{IV}(a). This organic
layer acts as a dielectric tunnel barrier preventing current to flow
between the electrodes and the nanowire. Irregularity in the
current/voltage characteristic occurs at current densities that are
large enough to physically destroy the surfactant layer. For current
above 5 to 10~mA, the typical resistance is below 100~$\Omega$
indicating a metallic electrical contact. This is demonstrated in
Fig.~\ref{IV}(b) where the current/voltage characteristic was
measured again during a subsequent bias sweep following the initial
run. A monotonous rise of the current with increasing voltage is
observed. For low voltage, the current/voltage characteristic
follows a linear trend (dashed curve) with a slope of 25~$\Omega$
corresponding to the total resistance of the contacts. The
resistance of the contacts between the probes and electrodes was
measured by contacting the two probes close to each other
(separation distance $<<$ the length of the nanowire) on the same
electrode. A real resistance of the Ag nanowire at around
2.35~$\Omega$ was thus determined yielding a resistivity $\rho$
=1.55~$\mu\Omega$.cm according to $\rho = (R \times S) / L$, where
$R$ is the subtracted resistance, $S$ is the cross-sectional area of
the silver nanowire, and $L$ is its longitudinal length. This
resistivity is very close to the standardized bulk value of silver
1.59~$\mu\Omega$.cm at room temperature~\cite{peng08}. After 0.3~V,
the trend becomes nonlinear with an increasing resistance of the
nanowire with voltage. This variation of resistance is due to the
temperature rise caused by the current flow~\cite{DurkanUM00} and
the onset of morphological changes induced by an electromigration
process~\cite{HuangIEEE08,StahlmeckeJPCM07,HoffmannAPL08}. The
effect of surfactant on the electron flow and the subsequent
monotonous current/voltage characteristics showed in
Fig.~\ref{IV}(a) and (b) were systematically measured for all
contacted Ag nanowires. The current/voltage characteristic displayed
in Fig.~\ref{IV}(b) is reproducible if the current flowing through
the nanowire is kept below the density required to initiate the
electromigration of Ag atoms. For voltage sweeps typically exceeding
1V, the subsequent on-start resistance increases as a result of
electromigration-induced amorphization of the nanowire.

\section{Surface plasmon analysis}

We now turn our attention to the properties of surface plasmons
propagating in these electrically-biased nanowires. Surface plasmons
were launched by using an oil immersion objective.  A laser emitting
at $\lambda=$800~nm was focused on a diffraction-limited spot at one
end of a contacted nanowire. Scattering at the other end of the
nanowire and leakage radiation during surface plasmon propagation
were collected by the same objective~\cite{SongACS11}. Two
charge-coupled device (CCD) cameras placed at the conjugated image
and Fourier planes of the microscope were used to record the
intensity distribution in direct and reciprocal space,
respectively~\cite{MassenotAPL07}. Measurements in the image plane
provide a direct visualization of plasmon propagation in the
nanowire while parameters like effective index of the guided surface
plasmon mode and overall losses were estimated from Fourier plane
analysis. Thicker nanowires were preferentially studied in this work
rather than nanowires with smaller sections since plasmon detection
by leakage radiation microscopy requires energy to leak into the
substrate~\cite{SongACS11}. Leakages only occur above a minimal
value of the section set by the local environment~\cite{zou10}.

\begin{figure}
\includegraphics[width=0.7\textwidth]{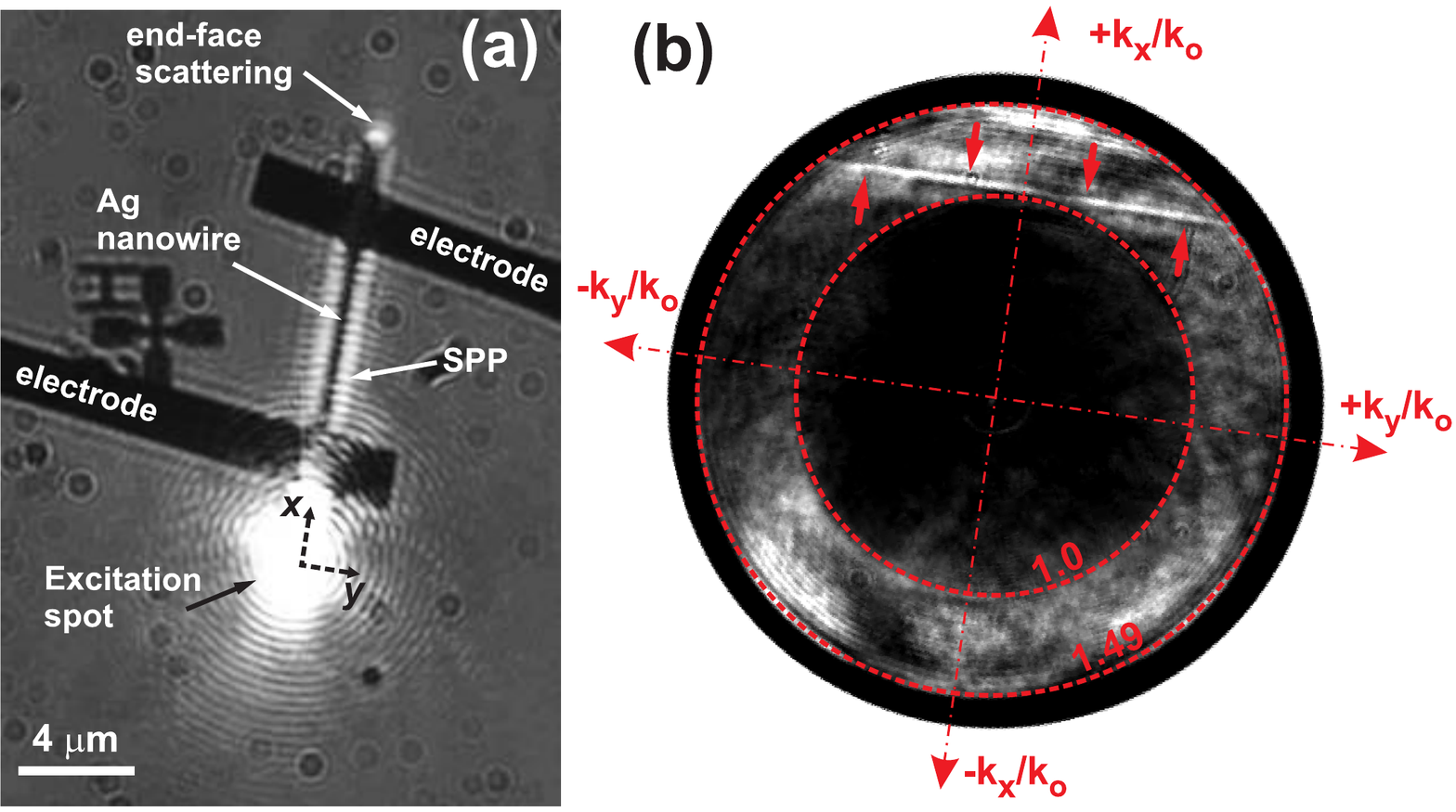}
\caption{(a) Leakage radiation image of a surface plasmon
propagating in an electrically contacted Ag nanowire. The excitation
laser spot polarized along the nanowire is adjusted with the lower
extremity. The reference frame is taken along the nanowire
($x$-axis). (b) Wavevector distribution obtained by Fourier-plane
imaging. The surface plasmon mode is recognized as a bright line at
a $+k_x/k_o$=1.056 (arrows). $|k_x/k_o|$= 1.49 and $|k_x/k_o|$=1.0
are given by the numerical aperture of the objective and the
critical angle at the glass/air interface, respectively.}
 \label{LK and FP}
\end{figure}

Figure~\ref{LK and FP}(a) displays a typical surface plasmon
intensity distribution recorded in the image plane for a contacted
Ag nanowire (section width 500~nm) excited with a linear
polarization oriented along the propagation. The electrodes are
visible in this image. The excitation spot overlapping the lower
extremity of the nanowire defines the origin of the reference
Cartesian frame ($x,y$), where $x$ is aligned with the long axis of
the nanowire. The surface plasmon mode is readily excited in this
configuration~\cite{LarocheAPL89} and leakages radiated during its
propagation are observed in the form of luminous lines on both sides
of the nanowire~\cite{Fang09NL,SongACS11}. At the distal end, the
surface plasmon is scattered out of the waveguide.

From Fig.~\ref{LK and FP}(a), there is no evidence that the plasmon
propagating along the uncoated sections of the wire is significantly
altered by the presence of the two 2-$\mu$m-wide electrodes. At the
electrode positions, the field distribution of the mode is modified
by the additional 70~nm-thick gold layer on top of the Ag nanowire
and no longer leaks in the substrate.  However, leakages resume on
the uncoated sections indicating that the mode is retained along the
pristine nanowire. We hypothesized that a significant field overlap
exists between the modes sustained at the electrode-coated sections
and the uncovered sections of the nanowire thus promoting energy
transfer between the modes~\cite{seidel03,seidel04}. This is
equivalent to a hybrid coupling strategies discussed in
Ref.~\cite{TongNL09}. A Fourier image of the wavevector distribution
is displayed in Fig.~\ref{LK and FP}(b). The leaky surface plasmon
mode is recognized as the unique bright line at an effective index
$N_{eff}=+k_x/k_o$=1.056 (arrows), where $k_x/k_o$ is the reciprocal
axis and $k_o=2\pi/\lambda$ is the free-space wavevector. The outer
boundary in Fig.~\ref{LK and FP}(b) indicates the collection limit
due to the objective numerical aperture $|k_x/k_o|$=1.49. The inner
boundary corresponds to critical angle of the glass/air interface at
$|k_x/k_o|$=1.0. The extension of the surface plasmon signature
along the $k_y/k_o$ reflects the lateral confinement of the mode on
the nanowire. The imaginary part of the complex effective index is
accounted by the width of the line and the measured value at
half-maximum $Im\{N_{eff}\}$=0.09 is comparable to uncontacted
nanowires~\cite{SongACS11}. Efficient scattering of the plasmon mode
at the distal end to free-space photons is typical for nanowires
with a section greater than $\sim$300~nm~\cite{kallOPEX12}.
Consequently, we did not observed a symmetric line at
$-k_x/k_o$=-1.056  which would suggest a reflected surface plasmon
mode by the end face.

\section{Simultaneous plasmon and electron transport}

\begin{figure}
\includegraphics[width=1\textwidth]{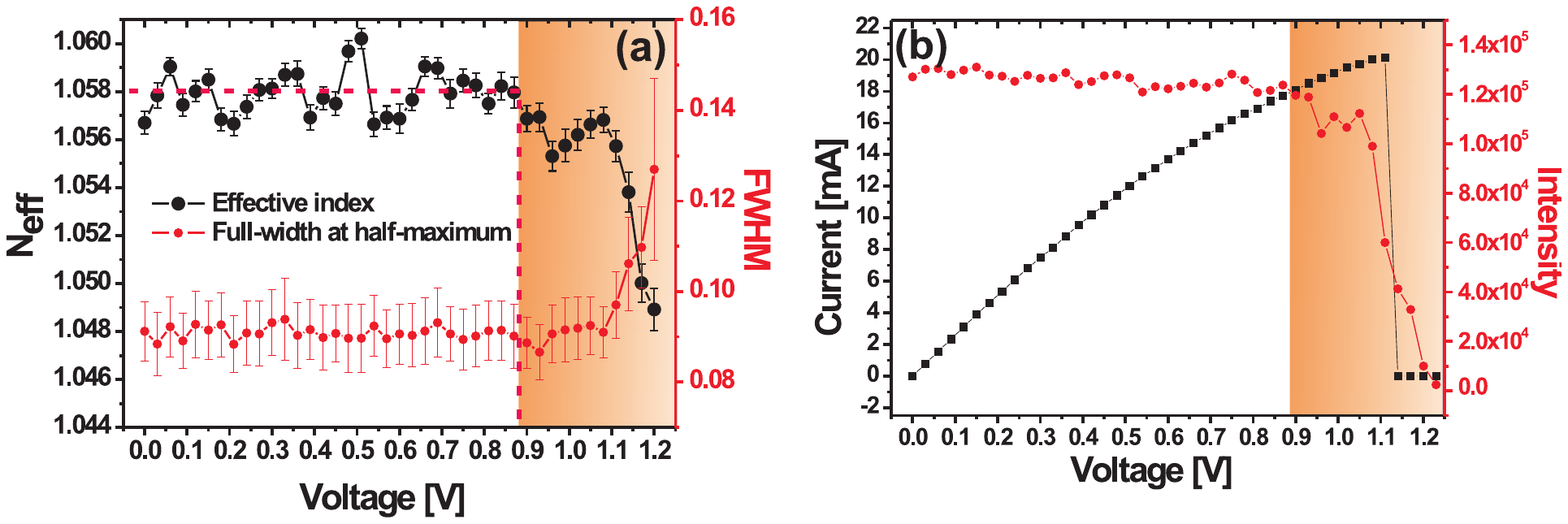}
  \caption{(a) Real and imaginary part of the surface plasmon
effective index in a Ag nanowire (section 500~nm) for increasing
biases. Both parameters are obtained by a Lorentzian fit of the
cross-section taken along the reciprocal $+k_x/k_o$ axis in
Fourier-plane measurement. The error bars are given by the fit
process. The dotted line represents the average value of
$Re\{N_{eff}\}$ in the voltage range [0;0.9~V]. (b) Data represented
in red circles: Evolution of intensity of light scattered at the end
of the nanowire with bias voltage. Data in black squares:
Simultaneously acquired current-voltage output characteristics of
the Ag nanowire. The sudden current drop at 1.1~V indicates
electrical breakdown of the nanowire.}
 \label{neff g5 nw10}
\end{figure}

Surface plasmon propagation was simultaneously recorded in this
dual-plane imaging mode while synchronously sweeping the DC bias at
30~mV steps. For each voltage/current values, a direct space image
and its corresponding Fourier content were recorded. We have then
extracted the value of the real and imaginary part of the effective
index for each biasing step. The results are reported in
Fig.~\ref{neff g5 nw10}(a).

The characteristics of the mode remain almost constant within
experimental measurement errors at $N_{eff}$=1.056+i0.09 until
0.9~V. The dotted line shows the average value of $Re\{N_{eff}\}$ in
the [0;0.9~V] voltage range. In this biasing regime, propagation of
the leaky plasmon mode is not affected by the flow of charges and
the temperature rise of the nanowire. Above 0.9~V (shaded areas),
$Re\{N_{eff}\}$ drops below the average value and both the real and
imaginary part of the complex effective index become dependent on
the biasing condition. Extrinsic propagation losses encoded in the
rise of $Im\{N_{eff}\}$ are concomitant to a measurable reduction of
$Re\{N_{eff}\}$. After 1.2~V, there is no longer any evidence of a
plasmon signature in the Fourier plane. The rise of propagation loss
after 0.9~V is confirmed by the rapid drop of the scattered
intensity with voltage measured at the distal end of the nanowire as
shown by the red curve (circle) in Fig.~\ref{neff g5 nw10}(b).

The change of the effective index and propagation loss is closely
related to the electrical output characteristic of the nanowire
reported again in Fig.~\ref{neff g5 nw10}(b). The plasmon
propagation is affected by the applied bias at the point where
electromigration of the nanowire becomes predominant in the
current/voltage characteristic (shaded area). The largest variation
of $Re\{N_{eff}\}$ and $Im\{N_{eff}\}$ occurs when the current
suddenly drops indicating electrical breakdown of the nanowire.
Under such electrical stress, the morphology of the nanowire is
significantly affected~\cite{XuJPCC10} thus compromising plasmon
propagation.

\begin{figure}
\includegraphics[width=0.8\textwidth]{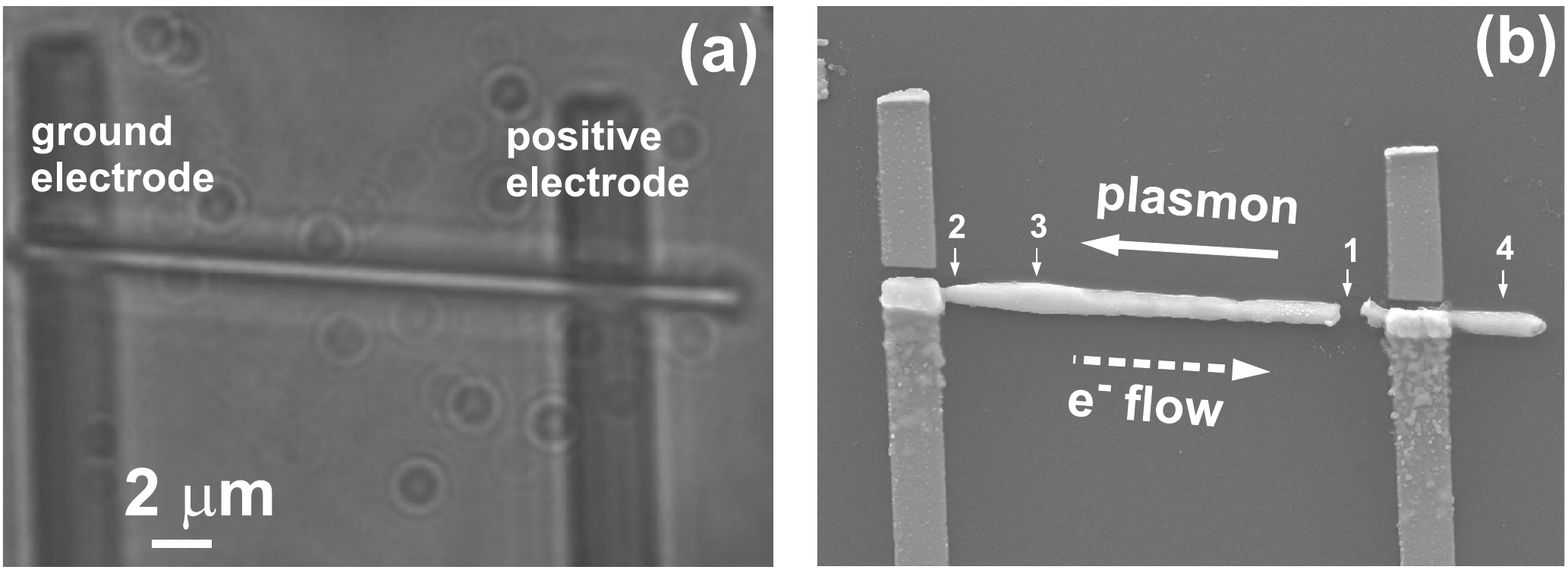}
\caption{ (a) Bright-field optical image of a 18~$\mu$m-long
500~nm-wide contacted Ag nanowire before applying bias. (b) Scanning
electron micrograph (SEM) of the same nanowire after electrical
breakdown. The surface plasmon excited at the positive leads is
propagating opposite to that of electron motion. The
electromigration causes electrical failure at the positive
electrode. The two gaps between the nanowire and the upper section
of the electrodes are caused by the angled evaporation of the metal.
Point 1 indicates location of the electrical breakdown, points 2 and
3 the morphological changes of the section due to mass transport,
and point 4 the pristine nanowire.}
 \label{bf and SEM images}
\end{figure}

To confirm this drastic morphological change, we have imaged the
nanowires before applying a bias and after electrical breakdown.
Freshly contacted nanowires were observed by bright-field optical
microscopy in order to avoid electron-beam carbon contamination by a
scanning electron microscope (SEM). After electromigration however,
a thin layer of Au ($<$20~nm) was sputtered on the sample for a
detailed SEM investigation. Representative images of a
18~$\mu$m-long 500~nm-wide Ag nanowire before and after
electromigration are shown in Fig.~\ref{bf and SEM images}(a) and
(b), respectively. The SEM image obtained after electromigrating the
nanowire confirms the structural deformation. For all investigated
Ag nanowires, electrical failure occurs at the positive electrode
(point 1). This behavior, already reported for another type of
crystalline Ag nanowires~\cite{StahlmeckeAPL06,kaspers09}, suggests
that the breakage originates from a motion of charged atoms
initiated by the electric field present across the nanowire.
However, Fig.~\ref{bf and SEM images}(b) also shows a weaker but
visible mass transport along the electron flow: a thinner nanowire
section near the ground electrode (point 2) is visible in the SEM
image. This section is immediately followed by a wider area (point
3) indicating that Ag atoms were displaced from point 2 to point 3.
The original nanowire section is measured at point 4, on the right
of the positive electrode. At this location, the electron flow did
not affect the nanowire. Note that the section at point 3 is wider
than the original nanowire size at point 4. We have also switched
the biasing polarity, i.e. electrons and plasmon propagating in the
same direction. We did not observe significant differences from the
trend displayed in Fig.~\ref{neff g5 nw10}(a).

\section{Numerical determination of the surface plasmon effective index}

\begin{figure}
\includegraphics[width=0.65\textwidth]{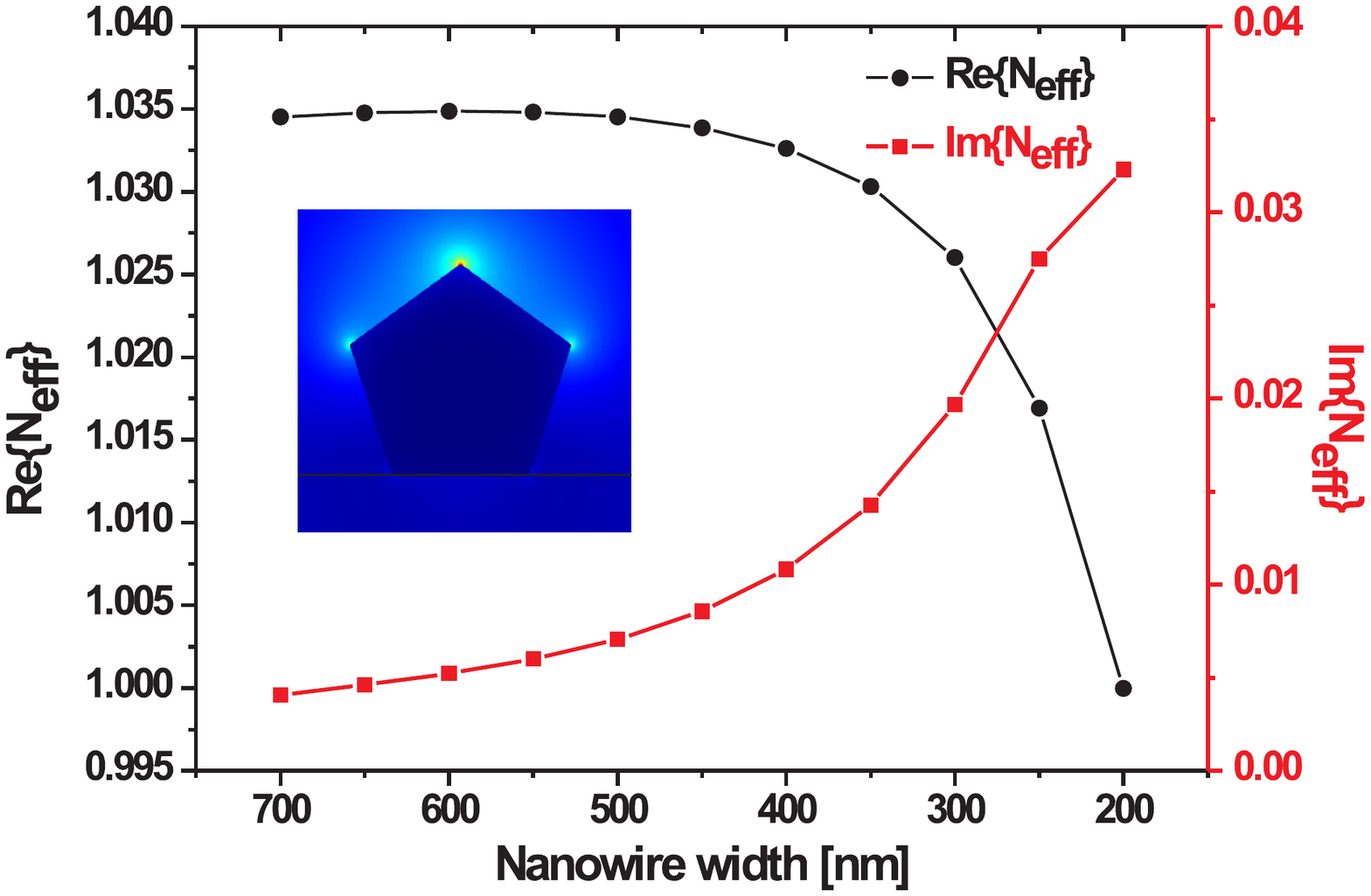}
  \caption{Calculated dispersion of the real and imaginary parts of the surface plasmon
effective index as a function of the nanowire section. The inset
shows a cross-sectional intensity distribution of the leaky plasmon
mode for a pentagonal Ag nanowire with a section of 300~nm.}
 \label{Dispersion curve}
\end{figure}

From Fig.~\ref{bf and SEM images}(b) it is clear that the evolution
of the surface plasmon effective index and propagation loss with
applied bias described in Fig.~\ref{neff g5 nw10}(a) can be
attributed to the altered section of the nanowire and loss of
crystallinity caused by the electromigration process. To test this
hypothesis, we numerically investigated the effect of a varying
nanowire section on the effective index and propagation loss of the
surface plasmon using the finite-element analysis simulation
software COMSOL Multiphysics. We considered an infinitely long
pentagonal silver nanowire on a glass substrate with an excitation
wavelength $\lambda = 800$~nm and a silver optical index $n_{Ag} =
0.0362+i5.4$. Since the effective index of the leaky plasmon mode is
very close to 1.0, the mode is significantly extending in the air
medium. To account for this, we add perfectly-matched layers on the
air side of the calculation window. The evolution of the effective
index with reducing nanowire section is reported in
Fig.~\ref{Dispersion curve}; the cross-sectional intensity
distribution of the mode is displayed in the inset for a section of
300~nm. The $Re\{N_{eff}\}$ and $Im\{N_{eff}\}$ are weakly depending
on nanowire section until approximatively 450~nm. For width
$<400$~nm, $Re\{N_{eff}\}$ monotonously decreases towards the cutoff
value of the mode (1.0), a condition confirmed by the associated
larger losses in this section range. A small discrepancy between the
experimental and calculated $Re\{N_{eff}\}$ can be attributed to the
dielectric constant used, the presence of the PVP polymer on the
surface of the nanowire, and experimental errors in calibrating
Fourier planes. A larger discrepancy is however present for the
$Im\{N_{eff}\}$. For a 500~nm-wide nanowire, we experimentally
measured $Im\{N_{eff}\}$=0.09 while the calculation indicates a
value approaching 0.01. Residual roughness present along the
chemically-synthesized Ag nanowire is probably contributing to the
widening of the Fourier signature of the surface plasmon mode. We
showed in Ref.~\cite{SongACS11} that the finite length of the
nanowire is affecting $Im\{N_{eff}\}$. End-face scattering
introduces an additional decay channel contributing to the total
loss experienced by the leaky surface plasmon mode.

According to SEM images of electromigrated nanowires, the thinnest
unbroken section is around 250~nm to 300~nm. The calculated
dispersion curve shows that the $Re\{N_{eff}\}$ of the mode is
reduced from 1.034 to $\sim1.020$, a change of 0.014 in this section
range. Although the relationship between the width of the nanowire
and the applied bias is not trivial (non linear electrical
characteristic and local variations along the length), the measured
difference of $Re\{N_{eff}\}$ before and after breakdown in
Fig.~\ref{neff g5 nw10}(a) ($\sim$0.010) is comparable to our
numerical simulations. Concerning the imaginary part
$Im\{N_{eff}\}$, the calculation also qualitatively reproduces the
experimental trend of Fig.~\ref{neff g5 nw10}(a). Amorphization of
the chemically-synthesized nanowire during the electromigration
process and variation of the section along the nanowire length were
omitted in the calculation. Both effects introduce additional
plasmon damping compared to pristine nanowires~\cite{KrennPRL05} and
accounts for the under approximation of the plasmon losses in
Fig.~\ref{Dispersion curve}.

\begin{figure}
\includegraphics[width=0.65\textwidth]{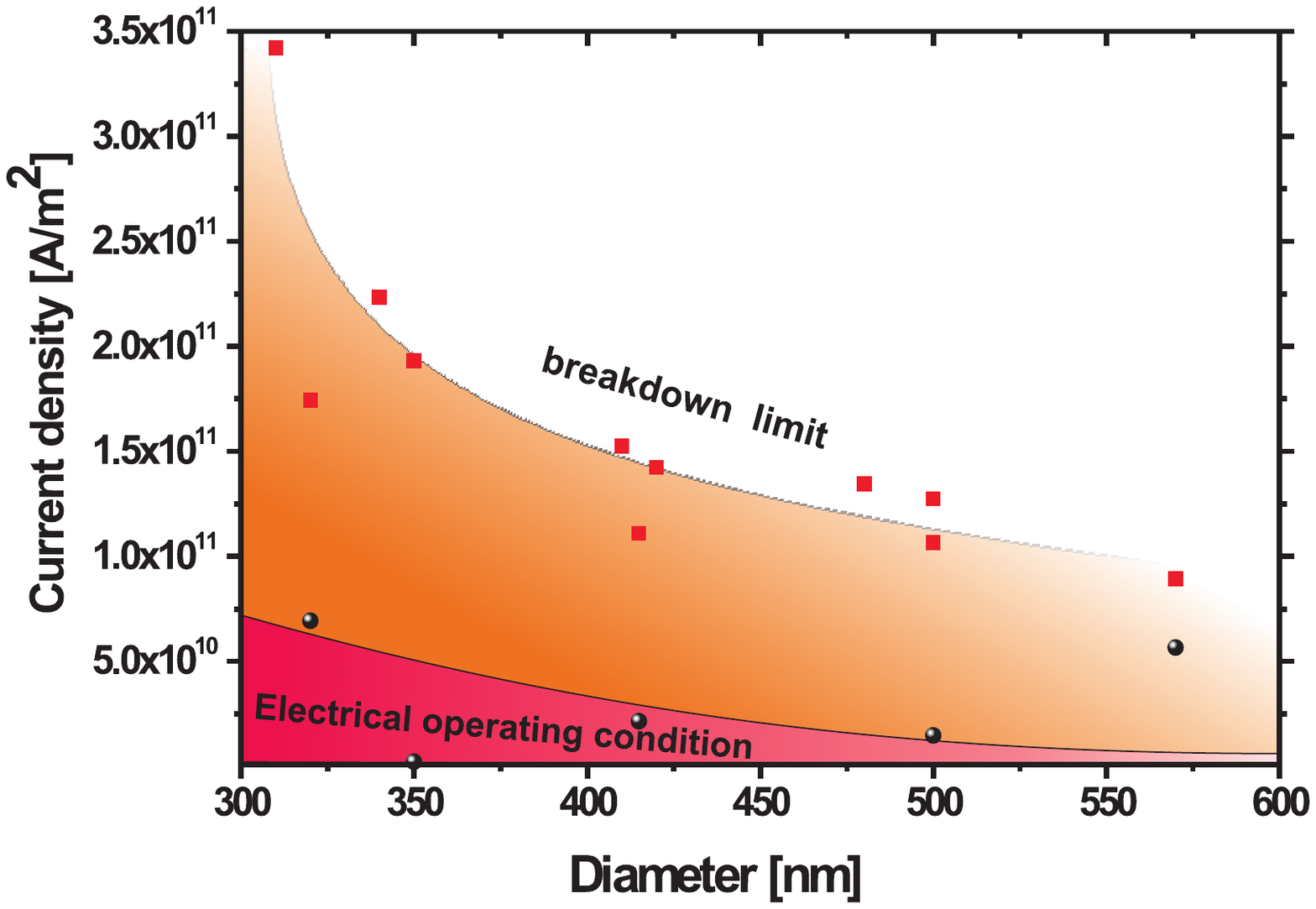}
  \caption{Current densities leading to electrical failure of crystalline Ag nanowires (squares) and degradation of
  surface plasmon characteristics (circles) as a function of nanowire width, respectively. The pink shaded area qualitatively
  indicates the safe electrical operating condition to simultaneously propagate
  electrons and surface plasmons in the metal nanowires.}
 \label{current density}
\end{figure}

From the above investigations, change in the morphology of the
nanowire appears to be the main limitation for co-propagating
plasmonic and electronic information in metal nanowires.
Figure~\ref{current density} shows the maximum current densities
before electrical failure (red squares) as a function of nanowire
section. Like Cu and Au nanowires, the failure current density in
crystalline Ag structures depends on the section as a results of a
better heat dissipation in the glass
substrate~\cite{DurkanUM00,HuangIEEE08,HuangJAPL08}. The data points
represented with black circles are the measured current densities
after which the corresponding $Re\{N_{eff}\}$ of the surface plasmon
mode drops below the average value measured at low voltage bias (see
Fig.~\ref{neff g5 nw10}(a)). This boundary is therefore an upper
limit on the the electrical condition under which electron and
(leaky) surface plasmon can be simultaneously supported by the
plasmonic platform. The red-shaded area beneath these data points
represents the safe electrical operation. In the orange area located
between the two curves, electron-induced degradation of the surface
plasmon mode is occurring. Thin nanowires can sustain a higher
current density and, from an electrical point of view, should be
favored in the circuitry. However, surface plasmons in thin
nanowires are tightly bound modes and are thus expected to be more
sensitive to subtle change occurring at the nanowire surface. We
hope to provide experimental evidence for that in a follow-up
communication.

In summary, we have investigated the effect of electron flow on
plasmon propagation in electrically contacted chemically-synthesized
silver nanowires. We find that surface plasmon characteristics are
degraded by the morphological stress of the nanowire caused at the
onset of electromigration. We have thus determined the operating
limit for simultaneously propagating electrons and surface plasmon
in the same one-dimensional metallic support, a prerequisite for
realizing a practical plasmonic circuitry interfacing integrated
photonic and electronic devices.

\ack

The research leading to these results has received funding from the
European Research Council under the European Community's Seventh
Framework Programme FP7/2007-2013 Grant Agreement no 306772 and
Grant ERC-2007-StG No. 203872-COMOSYEL. This work was also partially
funded by the Agence Nationale de la Recherche (ANR) under Grant
Plastips (ANR-09-BLAN-0049) and the R\'egion de Bourgogne under the
PARI program. M.S. acknowledges a stipend from the Chinese
Scholarship Council. D.Z. thanks support from National Natural
Science Foundation of China for Grants No. 11004182 and 61036005.

\end{document}